\documentclass[prl,twocolumn]{revtex4-2}
\usepackage{graphicx}
\usepackage{bm}
\usepackage{natbib}
\usepackage{xcolor}

\begin{document}
	
	\title{Collapse of orthotropic spherical shells}
	\author{Gautam Munglani\textsuperscript{1,2}}
	\affiliation{%
		\textsuperscript{1}Computational Physics for Engineering Materials, Institute for Building Materials, ETH Z\"urich, 
		Stefano-Franscini-Platz 3; CH-8093 Z\"urich, Switzerland
	}%
	\author{Falk K. Wittel\textsuperscript{1}}
	\affiliation{%
		\textsuperscript{1}Computational Physics for Engineering Materials, Institute for Building Materials, ETH Z\"urich, 
		Stefano-Franscini-Platz 3; CH-8093 Z\"urich, Switzerland
	}%
	\author{Roman Vetter\textsuperscript{1}}
	\affiliation{%
		\textsuperscript{1}Computational Physics for Engineering Materials, Institute for Building Materials, ETH Z\"urich, 
		Stefano-Franscini-Platz 3; CH-8093 Z\"urich, Switzerland
	}%
	
	\author{Filippo Bianchi\textsuperscript{1}}
	\affiliation{%
		\textsuperscript{1}Computational Physics for Engineering Materials, Institute for Building Materials, ETH Z\"urich, 
		Stefano-Franscini-Platz 3; CH-8093 Z\"urich, Switzerland
	}%
	\author{Hans J. Herrmann\textsuperscript{1,3}}
	\affiliation{%
		\textsuperscript{1}Computational Physics for Engineering Materials, Institute for Building Materials, ETH Z\"urich, 
		Stefano-Franscini-Platz 3; CH-8093 Z\"urich, Switzerland
	}%
	
	\author{}
	\affiliation{
		\textsuperscript{2}Institute of Plant and Microbial Biology, University of Z\"urich, Zollikerstrasse 107, CH-8008 Z\"urich, Switzerland
	}%
	\author{}
	\affiliation{
		\textsuperscript{3}Physique et M\'{e}canique des Milieux H\'{e}t\'{e}rog\`{e}nes (PMMH), ESPCI, 7 quai St.~Bernard, 75005 Paris, France
	}
	
	\begin{abstract}
		We report on the buckling and subsequent collapse of orthotropic elastic spherical shells under volume and pressure control. Going far beyond what is known for isotropic shells, a rich morphological phase space with three distinct regimes emerges upon variation of shell slenderness and degree of orthotropy. Our extensive numerical simulations are in agreement with experiments using fabricated polymer shells. The shell buckling pathways and corresponding strain energy evolution are shown to depend strongly on material orthotropy. We find surprisingly robust orthotropic structures with strong similarities to stomatocytes and tricolpate pollen grains, suggesting that the shape of several of Nature's collapsed shells could be understood from the viewpoint of material orthotropy.
	\end{abstract}
	
	\maketitle
	
	The collapse of spherical shells is as ubiquitous in Nature and technology as it is strikingly diverse, producing a myriad of post-buckled shapes when the delicate balance between geometry, material properties, and kinetics is perturbed. Understanding this behavior is necessary for a multitude of applications ranging from the prevention of catastrophic failure in water tanks and submarines \cite{Cole1969,Bushnell1981}, and the production of functional colloids \cite{Gao2001,Quilliet2008,Tsapis2005}, to ensuring the survival of pollen during dispersal by harmomegathy \cite{Katifori2010,Payne1972}, and predicting healthy human blood cell shapes \cite{Lim2002,Park2010}. Factors like the Poisson ratio \cite{Quilliet2012,Quemeneur2012}, transverse shear and effective bending stiffness \cite{Ru2009}, viscoelasticity \cite{Tsapis2005}, and compression rate \cite{Vliegenthart2011} have already been implicated in controlling the buckling pathways and post-buckling conformations for homogeneous spherical shells. Furthermore, inhomogeneities like variable cell wall thickness \cite{Datta2012}, local soft spots \cite{Paulose2013}, and spheroidal geometries \cite{Pauchard2004} exert an even stronger influence in directing collapse. Computer simulations, along with non-linear shell theory \cite{Quilliet2006,Knoche2011,Knoche2014} have proved invaluable in elucidating the collapse mechanics of isotropic spherical shells \cite{Vliegenthart2011,Quilliet2008,Quilliet2012}. However, shells with anisotropic properties, a common feature of composite materials like biological cell walls \cite{Baskin2005,Kerstens2001}, have also been simulated by applying identical isotropic assumptions, with mixed results \cite{Quemeneur2012}. So far, material anisotropy as a potential key player in shell morphogenesis has been largely excluded from thin shell analyses.
	\begin{figure}[htb]
	\centering
	\includegraphics[width=1\linewidth]{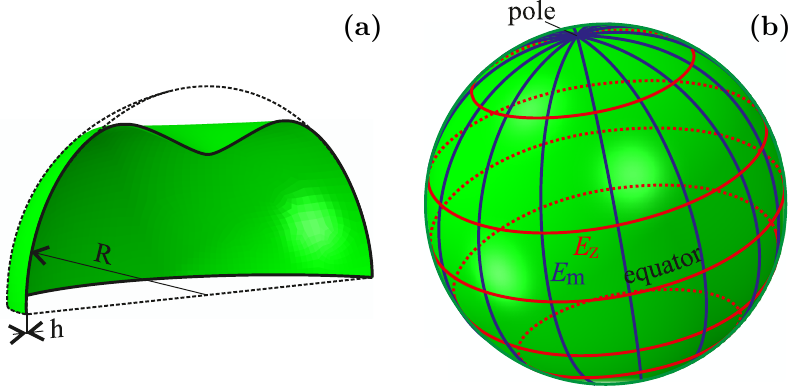}
	\caption{(a) Schematic representation of a single indentation on a spherical shell with radius $R$ and thickness $h$. (b) Illustration of the meridional and zonal directions on the spherical shell with their moduli $E_\mathrm{m}$, $E_\mathrm{z}$.}
	\label{fig:spheres}	
	\end{figure}
	\begin{figure*}[t]
	\centering
	\includegraphics[width=1\linewidth]{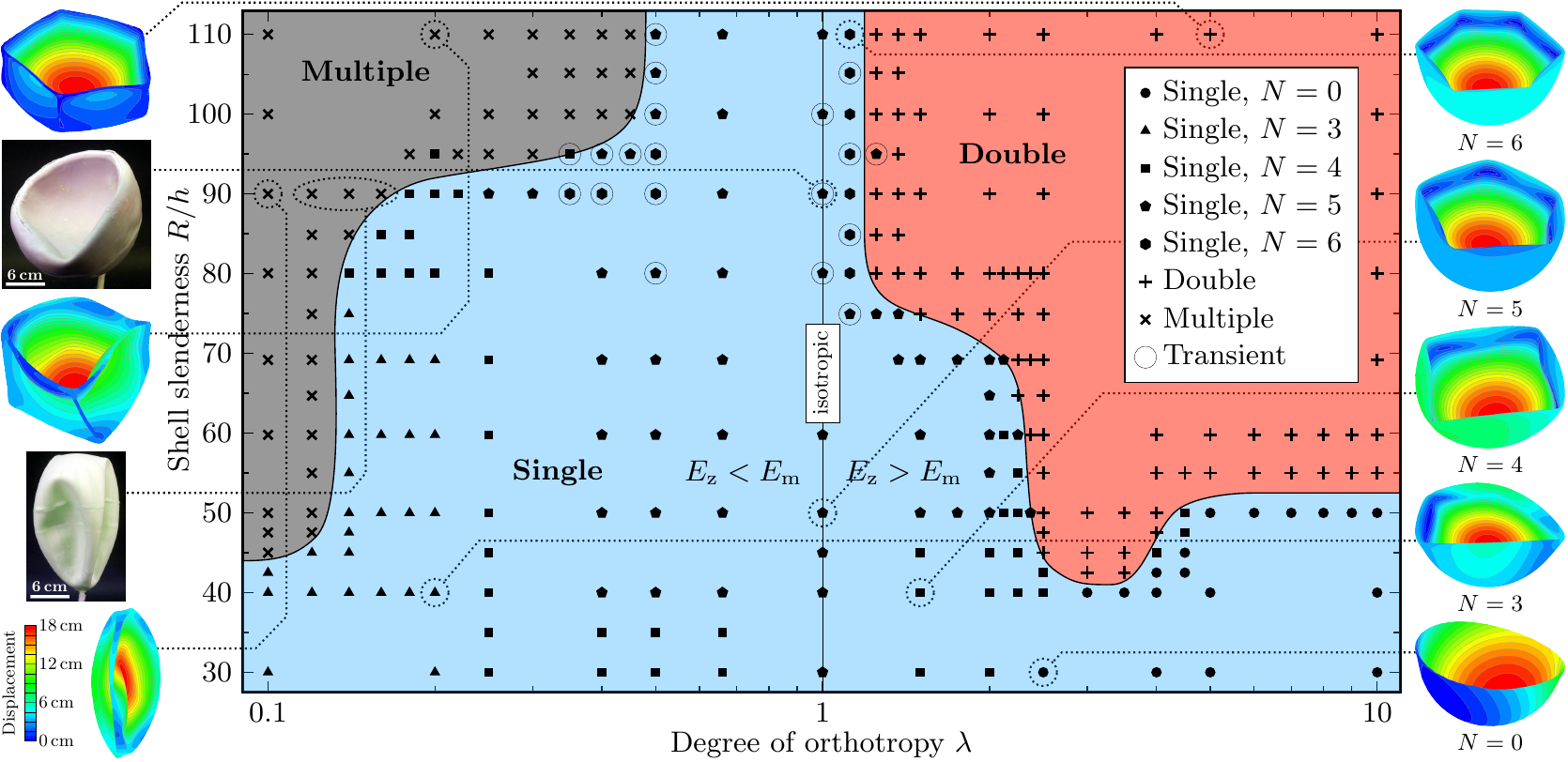}
	\caption{Post-buckling conformations of spheres with varying degree of orthotropy $\lambda$ and slenderness $R/h$ {($\Delta V/V_0 = 0.9$)}. Three distinct buckling phases are distinguishable -- single, biconcave double and multiple indentations. The shape of the marker in the single indentation phase represents the number of corners, $N$, of its polygonal ridge. Solid lines surrounding a marker indicate transient behavior, implying that multiple indentations formed before relaxing to a single indentation. The colors on the simulation figures correspond to radial displacement. {Movies of each of the three phases can be found in the Supplemental Material \cite{Supp}}. Experiments: Side view of collapsed shells ($R/h = 90$) floating underwater for orthotropic ($\lambda = 0.15$) and isotropic ($\lambda = 1$) systems.}
	\label{fig:buckle}	
	\end{figure*}

	In this work, a combination of numerical and experimental approaches are used to reveal the diversity of tunable and robust post-buckled conformations produced by the controlled collapse of in-plane orthotropic spherical shells. We find that adjusting the ratio of elastic properties between orthogonal directions and varying the slenderness of the shell alters the strain energy landscape of the collapse. This material irregularity therefore regulates the initial buckling and subsequent folding pathway, driving the system towards three distinct classes of conformations based largely on the degree of orthotropy. We find that there is a material regime in which orthotropic spherical shells can be more stable than isotropic ones under volume-controlled collapse. Finally, we present the scaling of the critical pressure at which buckling is initiated depending on these two control parameters. With this new insight into the stability of spherical shells, research on the morphogenesis of pollen grains, red blood cells, gel phase vesicles, blastulae, and related structures, is offered a new perspective in explaining the rich variety of observed shapes.
	
	Thin isotropic spherical shells  of radius $R$, wall thickness $h$, Young's modulus $E$, and Poisson ratio $\nu$, have been shown to collapse under a critical external pressure $p_\mathrm{ci}$ in a process resembling a first-order transition given by \cite{Zoelly1915,VonKarman1939}
	\begin{equation} \label{eq:cpressure}
	p_\mathrm{ci} = \frac{4E}{\sqrt{12(1-\nu^2})} \left(\frac{h}{R}\right)^2 \, .
	\end{equation}
	$p_\mathrm{ci}$ is inversely proportional to the square of the shell slenderness, which we define as $R/h$. When subjected to an increasing uniform external pressure $p<p_\mathrm{ci}$, the spherical shell contracts isotropically, resulting in increased elastic energy due to in-plane compression, while maintaining a constant bending energy. When $p_\mathrm{ci}$ is reached, the compressive stress is released at the cost of a transverse deflection in the form of a local indentation (Fig.~\ref{fig:spheres}(a)). In anisotropic shells, some areas are more prone to grow inwards than others, leading to symmetry breaking and a lower critical pressure, $p_\mathrm{c}$. 
	
	In Nature, buckled spheroidal shells are usually found as a result of a prolonged period of fluid loss, either by drying or controlled expulsion, rather than an immediate collapse at a constant external pressure \cite{Lim2002,Quilliet2008,Katifori2010,Park2010}. The controlled collapse of a spherical shell can be simulated numerically using a finite element solver coupling shell to hydrostatic fluid elements \cite{Abaqus2014,Vetter2013} (as opposed to mass-spring models published in earlier works \cite{Quilliet2008,Vliegenthart2011}). The interaction between the shell and the encapsulated fluid corresponds to a fluid cavity constraint, given by $V-\overline{V}=0$, where $\overline{V}$ is the volume enclosed by the shell and $V$ is the volume of the fluid. The fluid volume is controlled by making its fluid density $\rho_t$ a function of simulation time $t$, using the relation $\rho_t = (1 + \alpha t) \rho_0$, where $\alpha$ is the expansion coefficient and $\rho_0$ is the fluid density in the original unstressed configuration. Unlike previous studies on isotropic shells \cite{Quilliet2008,Quemeneur2012}, we limit ourselves to studying equilibrium post-buckling conformations by setting $\alpha=-0.1\,\mathrm{s}^{-1}$ and simulating $t=0\to10\,\mathrm{s}$. This low fluid volume compression rate ensures that our simulations traverse the energy landscape with a negligible probability of getting trapped in local minima.
	
	The most intuitive way to prescribe orthotropic properties on the surface of a spherical shell is with reference to its poles. The elastic modulus $E_\mathrm{m}$ applies in the meridional direction, while $E_\mathrm{z}$ is defined in the zonal direction parallel to the equator as seen in Fig.~\ref{fig:spheres}(b). The meridians intersect at the poles, while the zonal lines remain parallel. This geometric setup implies that different buckling regimes occur when the ratio of the elastic moduli $\lambda=E_\mathrm{z}/E_\mathrm{m}$ is greater or less than unity. {Details on the material properties and numerical implementation can be found in the Supplemental Material \cite{Supp}.}
	
	A variety of different post-buckled conformations are simulated in the parameter space of $R/h$ and $\lambda$ (Fig.~\ref{fig:buckle}). Three distinct phases of collapsed shells are identified - single, symmetric biconcave double, and multiple indentations. The final post-buckled shapes are always found to be in the single indentation phase in the isotropic case ($\lambda = 1$), in agreement with some previous studies on spherical shells with zero spontaneous curvature \cite{Quilliet2012,Vliegenthart2011}. In the isotropic case and for slender orthotropic shells ($R/h \approx 70$--$110$) close to the boundaries of the single indentation phase, the structures can sometimes retain up to three indentations even with a low reduced volume fraction, $\Delta V/V_0 \leq 0.9$ (where $V_0$ is the initial and $\Delta V$ is the lost fluid volume), { before eventually undergoing a merging process to form a single indentation, thereby showcasing transient behavior in the system \cite{Vliegenthart2011}.} {This low number of stable indentations (which we later confirm with our experiments) runs contrary to previous simulation studies for isotropic shells with much lower slenderness values ($R/h \le 20$) \cite{Quilliet2008,Quemeneur2012}, which were based on a reduced Helfrich energy functional that does not account for Gaussian curvature rigidity.}
	
	As $\lambda$ deviates from unity, the post-buckling shapes become highly dependent on the slenderness. For thicker shells ($R/h \approx 30$), the post-buckling conformations are still in the single indentation phase, but as the shells become thinner, two novel and distinct phases emerge. For $\lambda \approx 1.25$--$10$ and $70 \leq R/h \leq 110$, two indentations are initiated at opposite poles and grow inward at roughly the same rate until self-contact is realized. This is explained by the fact that the poles are converging points for meridional lines of lower modulus ($E_\mathrm{m}$), which gives them a lower bending stiffness, making them the areas on the shell with the least resistance against indentations. This biconcave shape is characteristic of healthy red blood cells \cite{ParkY2010}.
	
	For slender shells with $\lambda \approx 0.1$--$0.5$, post-buckling conformations exhibit multiple indentations. In this phase, the shells initially morph into prolate spheroids with progressively growing ellipticity under fluid loss. This observation can be explained by the zonal lines having a lower $E_\mathrm{z}$ than the poles, resulting in the compression of the spherical shell along the equator and other zonal lines. Multiple spherical cap inversions then form, grow inwards, and remain robust even at very high $\Delta V/V_0$. The poles are the most stable areas on the spherical shell due to the convergence of the stiffer meridional lines, and the mean curvature $\kappa_\mathrm{m}$ is highest at the poles and lowest in the equatorial region. As the geometry-induced stiffness has been shown to be proportional to $\kappa_\mathrm{m}$ in spheroidal shells \cite{Lazarus2012}, indentations are expected to initiate around the flexible equatorial region. For highly orthotropic shells, with $\lambda \approx 0.1$ and $R/h = 50$--$90$, the simulations follow this prediction, as a large initial change in the ellipticity drives the shell towards an elongated structure with 3--4 indentations along the meridional axis. This showcases a buckling pattern found in dehydrated tricolpate \textit{Brassica rapa} pollen grains, whose folding has been shown to be guided by the composite intine of the cell wall  \cite{Ferguson1998,Kirkpatricka2013}. However, for highly slender shells, with $R/h \approx 110$ and $\lambda \approx 0.25$, the change in ellipticity is much less significant, resulting in a very slight change in $\kappa_\mathrm{m}$ along the meridians. {No correlation between patterns of equivalent stress and locations of the onset of indentations could be observed prior to buckling}. This results in a cuboid-like structure with 5--6 indentations, similar to anisotropic gel phase vesicles in Ref.~\cite{Quemeneur2012}. {Given a buckled elastic homogeneous shell with a known thickness and diameter, one can infer the type of orthotropy ($\lambda < 1$ or $\lambda > 1$) from the placement and number of indentations.
	
	Secondary buckling modes also provide clues to $\lambda$, and are seen across the parameter space, generally along the periphery and perpendicular to the ridge of the initial indentation.} These secondary effects give the ridge an equilateral polygonal shape with a number of corners, $N$, that increases with the slenderness due to the associated reduction in bending stiffness. Shells with $\lambda \ll 1$ usually have triangular-shaped ridges as the greater ellipticity of the spheroid makes it less energy efficient to form a square or other higher order polygons, unless the shell is very slender. A similar triangular-shaped ridge is found to be present in stomatocytes, which are formed due to defects in the membrane bilayer of red blood cells, which has been linked to membrane orthotropy \cite{Fischer1981,Lim2002}. For $\lambda \gg 1$, low $R/h$ values lead to the absence of any secondary buckling modes, resulting in a perfectly smooth spherical cap inversion, similar to thick isotropic shells in Ref.~\cite{Quilliet2012}. For $0.25 < \lambda < 4$, $N$ generally increases with $R/h$ from $4 \rightarrow 5 \rightarrow 6$, with the exception of $\lambda = 1$, which remains pentagonal in the simulated range $30 < R/h < 110$. 
		\begin{figure}[htb]
		\centering
		\includegraphics{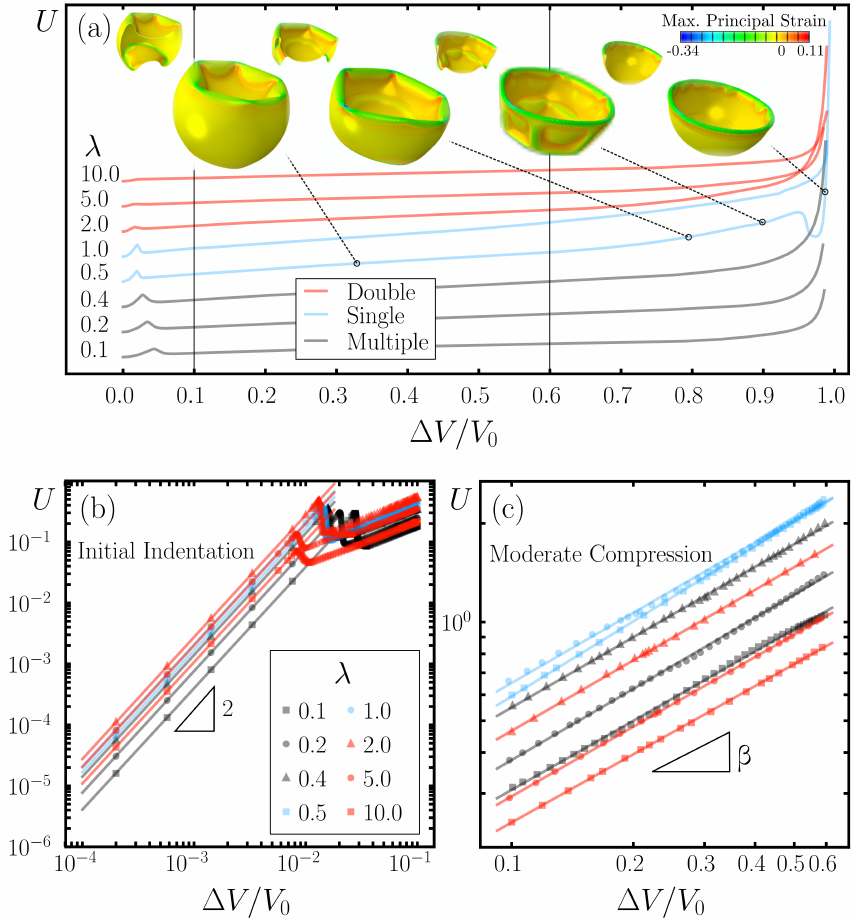}
		\caption{(a) Evolution of the strain energy $U$ (Joule) for different $\lambda$ at fixed slenderness $R/h=110$. The colors represent the single, biconcave double and multiple indentation phases from Fig.~\ref{fig:buckle}. Note that curves are vertically shifted for better visibility. Snapshots and their cuts are shown for the transient buckling case ($\lambda=0.5$) at specific reduced volume fractions. (b) $U$ scales quadratically with $\Delta V/V_0$ before the onset of the first indentation. (c) $U$ scales with $\Delta V/V_0$ as a power law with exponent $\beta=0.72\pm0.02$ in the region $0.1 \leq \Delta V/V_0 \leq 0.6$.}
		\label{fig:strainenergy}	
	\end{figure}
	\begin{figure}[t]
		\centering
		\includegraphics[width=\columnwidth]{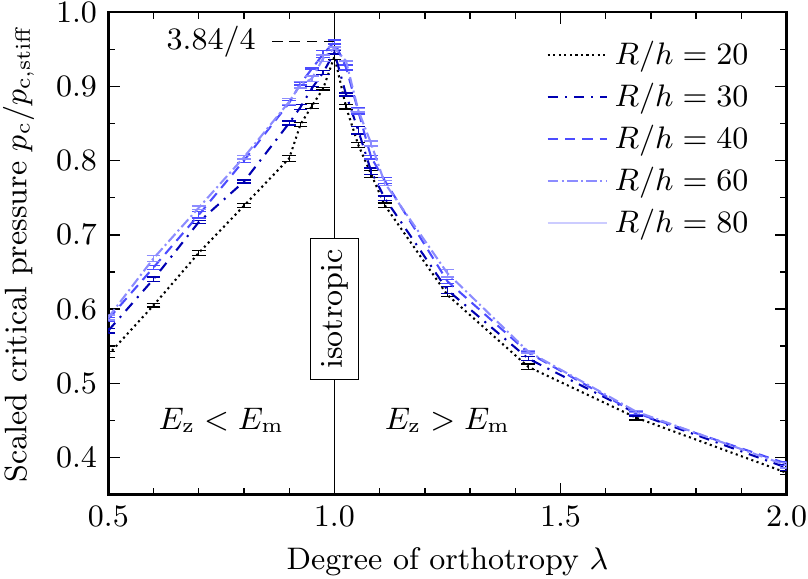}
		\caption{Scaling of the critical buckling pressure with $\lambda$ and $R/h$. Each data point represents the average over five independent simulations; error bars denote statistical errors.}
		\label{fig:pressure}
	\end{figure}
	The evolution of the strain energy $U$ gives further insight into shell collapse dynamics. At very high reduced volume fractions ($\Delta V/V_0 > 0.9$), $U$ is shown to increase significantly due to self-contact between the indentation(s) (Fig.~\ref{fig:strainenergy}(a)). The parameters $\lambda = 0.5$ and $R/h = 110$ represent the transient case, where for a majority of simulation runs, a large drop in $U$ is seen due to a reduction in the number of indentations at high $\Delta V/V_0$. For moderate volume fractions ($0.1 < \Delta V/V_0 < 0.6$), depending on $\lambda$, we find that $U\sim(\Delta V/V_0)^\beta$ with an exponent of $\beta=0.72\pm0.02$ (Fig.~\ref{fig:strainenergy}(b)), which is very close to previous work on isotropic shells \cite{Vliegenthart2011}. The strain energy in the region $\Delta V/V_0 = 0.1 \rightarrow 0.6$ is always largest at $\lambda = 1$ and reduces as $\lambda$ deviates from 1. $\lambda < 0.5$ has higher $U$ than $\lambda > 2.0$. This is expected, as $U$ has been shown to scale with the number of indentations with an exponent of 0.25 \cite{Quilliet2006}. Regardless of the shell properties, at $\Delta V/V_0 < 0.1$, there is a peak in $U$ corresponding to the initial spherical cap inversion. This increase in $U$ scales quadratically with $\Delta V/V_0$ (Fig.~\ref{fig:strainenergy}(c)). At $\lambda = 1$, the peak occurs at a relatively high $\Delta V/V_0$ compared to moderately orthotropic materials ($0.5 < \lambda < 2.0$), due to the inherent stability of the homogeneous isotropic shell. As $\lambda$ deviates further from unity, the peak shifts back to larger $\Delta V/V_0$ values, and in the case of $\lambda = 0.1$--$0.4$, even exceeds that of the isotropic case. While this might seem counter-intuitive, it can be explained by the effect of the geometric transformation of the shell towards ellipticity outweighing the loss in bending stiffness caused by the high degree of orthotropy. This effect is more pronounced for $\lambda < 0.5$, as these structures undergo larger geometric changes than for $\lambda > 2.0$, allowing the $\lambda < 0.5$ shells to withstand higher compressive stresses without collapsing.
	
	To support our numerical findings with experimental evidence, we reproduced the volume-controlled collapse with a versatile technique for fabricating shells \cite{Lee2016,Roman2018}. The procedure consists of pouring a silicone-based liquid polymer solution onto a steel sphere, resulting in hemispherical shells of relatively predictable thickness after the uncured polymer has drained. To create a spherical shell, two hemispherical caps ($R$ of $110$mm and $h$ of $1.2\pm0.06$\,mm) are produced separately and glued together at the equator (Fig. S1). In the orthotropic case, strips of thin PTFE film are glued on in the meridional direction to provide fiber reinforcement. The effect of the stiffer PTFE results in an orthotropic shell with a $\lambda$ of 0.15 using equivalent stiffness models for composites \cite{Xia2012}. These spherical shells are then filled with water through a valve and placed inside a water bath. In the case of the isotropic shell, ejecting the encapsulated water at different rates (over $1$--$100\,\mathrm{s}$) always showcases a robust single indentation, with $N = 4$--$5$ (Fig.~\ref{fig:buckle}). The orthotropic case yields a shape very similar to the elongated structure (similar to tricolpate \textit{Brassica rapa} pollen) found in the simulated multiple indentations phase.
	
	Studying the change in critical pressure $p_\mathrm{c}$ for shells with varying $R/h$ and $\lambda$ provides a quantitative description of their structural stability \cite{Vliegenthart2011}. Since orthotropic extensions of Eq.~\ref{eq:cpressure} do not exist for spherical or ellipsoidal shells to the best of our knowledge \cite{Krivoshapko2007}, our analysis of $p_\mathrm{c}$ for anisotropic shells can only be validated compared to the theoretical $p_\mathrm{ci}$. To simulate rapid collapse, subdivision surface shell elements within a dynamic time integration framework \cite{Cirak2000,Vetter2013,Vetter2014,Munglani2015} are used to ensure a high degree of accuracy.
	
	The numerical results confirm that $p_\mathrm{c}$  of the orthotropic spherical shells remains approximately proportional to $(R/h)^{-2}$ as in Eq.~\ref{eq:cpressure} as shown in Fig.~\ref{fig:pressure}. All pressure data approximately collapse onto a master curve after scaling by
	\begin{equation}
	p_\mathrm{c,stiff}=\frac{4\max\{E_\mathrm{m},E_\mathrm{z}\}}{\sqrt{12(1-\min\{\nu_\mathrm{zm},\nu_\mathrm{mz}\}^2)}}\left(\frac{h}{R}\right)^2 \, .
	\end{equation}
	The collapse converges in the membrane limit $R/h\to\infty$. $p_\mathrm{c}$ is scaled by $p_\mathrm{c,stiff}$ because in the orthotropic case, buckling can only be initiated when the bending energy barrier in both principal tangential directions is exceeded, making the critical pressure governed by the stiffer of the two in first order. This is consistent with the behavior of other types of material tuning \cite{Paulose2013}.
	
	The dependence of $p_\mathrm{c}$ on $\lambda$ reveals a distinct difference between the two orthotropic regimes $\lambda<1$ and $\lambda>1$. The theoretical prefactor of 4 ($3.84 \pm 0.03$ in the simulations) decreases as $\lambda$ deviates from 1. The regime with $\lambda < 1$ always provides larger prefactors than $\lambda > 1$ for the same logarithmic distance $\left|{\log\lambda}\right|$, as is expected from the controlled collapse analysis. While for $\lambda<1$ the critical pressure drops nearly linearly with the degree of orthotropy for sufficiently thin shells in the examined range, the scaling is nonlinear for $\lambda>1$ and does not appear to follow a simple functional relationship. A part of the observed nonlinearity in $\lambda$ stems from the nonlinear deformation of orthotropic spheres into spheroids under compression \cite{Krivoshapko2007}.
	
	Our study shows that material orthotropy and shell slenderness can be tuned to achieve three distinct regimes of post-buckled shapes. This suggests that the shapes of red blood cells and \textit{Brassica rapa} pollen occurring in Nature can now be examined through the lens of anisotropic properties, yielding insight into their structure-function relationship \cite{Katifori2010,Lim2002,Park2010}. We also find that strongly orthotropic materials can be more stable under controlled collapse than isotropic structures, despite having lower equivalent stiffness, resulting in potential applications in the fabrication of functional colloids. Finally, the variability of the scaling behavior of the critical pressure with the type of orthotropy ($\lambda < 1$ or $\lambda > 1$) poses an interesting query that requires further theoretical insight.
	
	\begin{acknowledgments}
	The authors would like to thank Benoit Roman for his assistance in shell fabrication and Alessandro Leonardi for helpful suggestions. Furthermore, the authors acknowledge support from the Research and Technology Development Project “MecanX: Physics-Based Models of Growing Plant Cells using Multi-Scale Sensor Feedback” granted under SystemsX.ch by the Swiss National Science Foundation, the Advanced Grant 319968-FlowCCS granted by the European Research Council (ERC) and from ETH Z\"urich by ETHIIRA Grant No. ETH-03 10-3. HJH would like to thank CAPES and FUNCAP.
	\end{acknowledgments}

	\nocite{Budiansky1963}
	\nocite{Li2014}
	\nocite{Huber1923}
	\nocite{Mang1980}
	\nocite{Oden1980}
	\bibliographystyle{apsrev4-2}
	\bibliography{main}

\end{document}

\title{Supplemental Material for Collapse of orthotropic spherical shells}
	\author{Gautam Munglani\textsuperscript{1,2}}
	\affiliation{%
		\textsuperscript{1}Computational Physics for Engineering Materials, Institute for Building Materials, ETH Z\"urich, 
		Stefano-Franscini-Platz 3; CH-8093 Z\"urich, Switzerland
	}%
	\author{Falk K. Wittel\textsuperscript{1}}
	\affiliation{%
		\textsuperscript{1}Computational Physics for Engineering Materials, Institute for Building Materials, ETH Z\"urich, 
		Stefano-Franscini-Platz 3; CH-8093 Z\"urich, Switzerland
	}%
	\author{Roman Vetter\textsuperscript{1}}
	\affiliation{%
		\textsuperscript{1}Computational Physics for Engineering Materials, Institute for Building Materials, ETH Z\"urich, 
		Stefano-Franscini-Platz 3; CH-8093 Z\"urich, Switzerland
	}%
	
	\author{Filippo Bianchi\textsuperscript{1}}
	\affiliation{%
		\textsuperscript{1}Computational Physics for Engineering Materials, Institute for Building Materials, ETH Z\"urich, 
		Stefano-Franscini-Platz 3; CH-8093 Z\"urich, Switzerland
	}%
	\author{Hans J. Herrmann\textsuperscript{1,3}}
	\affiliation{%
		\textsuperscript{1}Computational Physics for Engineering Materials, Institute for Building Materials, ETH Z\"urich, 
		Stefano-Franscini-Platz 3; CH-8093 Z\"urich, Switzerland
	}%
	
	\author{}
	\affiliation{
		\textsuperscript{2}Institute of Plant and Microbial Biology, University of Z\"urich, Zollikerstrasse 107, CH-8008 Z\"urich, Switzerland
	}%
	\author{}
	\affiliation{
		\textsuperscript{3}Physique et M\'{e}canique des Milieux H\'{e}t\'{e}rog\`{e}nes (PMMH), ESPCI, 7 quai St.~Bernard, 75005 Paris, France
	}
	\maketitle
	{
	\section{Non-linear finite element implementation}
	The spherical shell in the collapse simulations is represented by shear-flexible, large-rotation (Reissner--Mindlin) elements. The expression for the internal virtual work in this formulation reads  
	\begin{eqnarray}
	\delta W^\mathrm{int} &=& \int_{\Omega} \left(\delta \epsilon_{\alpha\beta} \int_{h} \sigma^{\alpha\beta} \, dz + \delta \kappa_{\alpha\beta} \int_{h} z \, \sigma^{\alpha\beta} \, dz \right) d\Omega \nonumber\\*
	&+& \, \sum_{i} K^{i}_{3\alpha} \, \gamma^{i}_{3\alpha} \, \delta \gamma^{i}_{3\alpha} \, ,
	\end{eqnarray}
	where $\delta \epsilon_{\alpha\beta}$ and $\delta \kappa_{\alpha\beta}$ are the first variations of the membrane and bending strain tensors over the middle surface $\Omega$, $\sigma^{\alpha\beta}$ is the Cauchy stress tensor on a surface offset by a distance $z$ from the middle surface over the shell thickness $h$, and $\gamma_{3\alpha}$ is the transverse shear strain with stiffness $K^{i}_{3\alpha}$ at a reduced set of integration points $i$. This stiffness is given by \cite{Abaqus2014}
	\begin{equation}
	K_{3\alpha} = \frac{G_{3\alpha} \, h \, \Delta\mathrm{A}}{(1 + q \, \Delta\mathrm{A}/h^2)} \, ,
	\end{equation}
	where $G_{3\alpha}$ are the elastic moduli associated with transverse shear, $\Delta\mathrm{A}$ is the reference surface area associated with $i$, and $q=0.25\times10^{-4}$ is a numerical factor. 
	
	The strain measure in this formulation corresponds to the Koiter--Sanders theory \cite{Budiansky1963}. Furthermore, the Cauchy stress tensor at a given $z$ is given by
	\begin{equation}
	\sigma^{\alpha\beta} = D_{\alpha\beta\gamma\delta} \, (\epsilon_{\gamma\delta} - z \, \kappa_{\delta\gamma}) \, ,
	\end{equation}
	where the plane stress elastic stiffness tensor, $D_{\alpha\beta\gamma\delta}$, defines the orthotropic material properties of the shell. The study of anisotropy requires careful consideration of Poisson ratio and shear modulus values to ensure physically meaningful results \cite{Quemeneur2012} and material stability given by a positive-definite elasticity tensor \cite{Li2014}. To maintain consistency with our experiments, we choose $\nu_\mathrm{zm}=\nu\sqrt{\lambda}$ and $\nu_\mathrm{mz}=\nu/\sqrt{\lambda}$ with fixed $\max\{\nu_\mathrm{zm},\nu_\mathrm{mz}\}=0.3$. In the pressure-controlled collapse simulations, we fix $\nu=0.3$. The shear modulus is set to $G_\mathrm{zm} = \frac{\sqrt{E_\mathrm{z}E_\mathrm{m}}}{2(1+\sqrt{\nu_\mathrm{zm}\nu_\mathrm{mz}})}$ \cite{Huber1923}, which uses the geometric mean of the Young's moduli and Poisson ratios.
	
	The pressure load stiffness arising from the cavity constraint results in the external virtual work given by
	\begin{equation}
	\delta W^\mathrm{ext} = \int_{\Omega} -p \delta V -\delta p (V-\overline{V})  \, d\Omega \, , 
	\end{equation}
	where $p$ is the pressure, $\overline{V}$ is the volume enclosed by the shell and $V$ is the actual volume of the fluid \cite{Mang1980}. For the volume-controlled simulations, $p$ takes the role of a Lagrange multiplier for the volume constraint.
	
	The finite element in this formulation contains 6 degrees of freedom (including 1 constrained rotational degree of freedom) per node and uses bi-quadratic interpolation functions for the displacements and rotations. Material randomness is introduced by multiplying the Young's moduli of individual elements by a random variable $X$ distributed as $\log(X) \sim \mathcal{N}(0,\sigma^2)$ with a standard deviation of $\sigma=0.001$. 
	
	Taking the long time scale of the volume compression process into account, a quasi-static system with appropriate damping is solved using an implicit, adaptive, backward Euler time-integration scheme. Thickness integration is performed using Simpson's rule. 
	
	Self contact of the shell during collapse is implemented with a finite-sliding, surface-to-surface discretization using the penalty method for constraint enforcement. In this formulation, the inner membrane of the cavity is defined as both the master and slave surface for finding contact pairs \cite{Oden1980}.}
	
	\section{Fabrication of orthotropic spherical shells}
	\begin{figure}[hb!]
		\centering
		\includegraphics[width=0.8\linewidth]{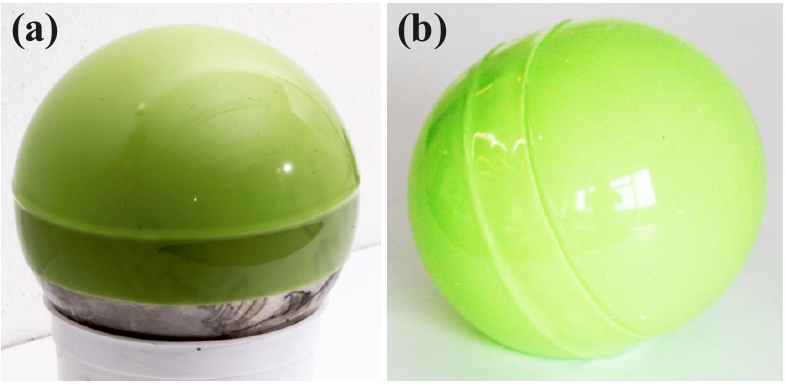}
		\caption{(a) Hemispherical cap with four layers on a sphere of radius 110\,mm. (b) Complete spherical shell after curing.}
		\label{fig:supplementary}	
	\end{figure}
	
	\bibliographystyle{apsrev4-2}
	\bibliography{main}